\documentclass{raa}

\usepackage{epsfig}
\usepackage{lscape,graphicx}
\usepackage{url}
\usepackage{enumerate}

\usepackage{graphicx,times}             
\input{epsf.sty}                        

\begin{document}

\title{An Algorithm for Preferential Selection of Spectroscopic Targets in LEGUE}


   \volnopage{Vol.0 (200x) No.0, 000--000}      
   \setcounter{page}{1}          

   \author{Jeffrey~L.~Carlin
      \inst{1}
   \and S\'ebastien L\'epine
      \inst{2,3}
   \and Heidi~Jo~Newberg
      \inst{1}
    \and Licai Deng
      \inst{4}
     \and Timothy C. Beers
       \inst{5}
    \and Yuqin Chen
      \inst{4}
    \and Norbert Christlieb 
      \inst{6}
    \and Xiaoting Fu
      \inst{4}
    \and Shuang Gao
      \inst{4}
     \and Carl J. Grillmair
       \inst{7}
    \and Puragra Guhathakurta
       \inst{8}
    \and Zhanwen Han
        \inst{9}
    \and Jinliang Hou
      \inst{10}
    \and Hsu-Tai Lee
      \inst{11}
    \and Jing Li
      \inst{4}
    \and Chao Liu
      \inst{4}
    \and Xiaowei Liu 
      \inst{12}
    \and Kaike Pan 
      \inst{13}
    \and J.~A. Sellwood
      \inst{14}
    \and Hongchi Wang
      \inst{15}
    \and Fan Yang
       \inst{4}
     \and Brian Yanny
       \inst{16}
    \and Yueyang Zhang
      \inst{4}
    \and Zheng Zheng
      \inst{17}
    \and Zi Zhu
      \inst{18} 
   }

   \institute{Department of Physics, Applied Physics, and Astronomy,
     Rensselaer Polytechnic Institute, 110 8th Street, Troy, NY 12180,
     USA (carlij@rpi.edu)\\
        \and
             Department of Astrophysics, American Museum of Natural
             History, Central Park West at 79th Street, New York, NY
             10024, USA\\
        \and
             The City University of New York, New York, NY, USA\\
         \and
             National Astronomical Observatories, CAS, Beijing 100012, China (NAOC)\\
        \and
            National Optical Astronomy Observatory, Tucson, AZ 85719, USA\\
        \and 
             Center for Astronomy, University of Heidelberg, Landessternwarte, Kšnigstuhl 12, D-69117 Heidelberg, Germany\\
         \and
            Spitzer Science Center, 1200 E. California Blvd., Pasadena, CA 91125, USA\\
        \and
            UCO/Lick Observatory, Department of Astronomy and Astrophysics, University of California, Santa Cruz, CA 95064\\
        \and
             Yunnan Astronomical Observatory, CAS, Kunming, China (YNAO)\\
        \and
            Shanghai Astronomical Observatory, CAS, Shanghai, China (SHAO)\\
        \and
            Academia Sinica Institute of Astronomy and Astrophysics, Taipei, China (ASIAA)\\
        \and
            Department of Astronomy \& Kavli Institute of Astronomy and Astrophysics, Peking University, Beijing 100875, China\\
        \and
            Apache Point Observatory, PO Box 59, Sunspot, NM 88349, USA\\
        \and
            Department of Physics and Astronomy, Rutgers University, 136 Frelinghuysen Road, Piscataway, NJ 08854-8019, USA\\
        \and
            Purple Mountain Observatory, CAS, Nanjing 210008, China (PMO)\\
        \and 
            Fermi National Accelerator Laboratory, P.O. Box 500, Batavia, IL 60510, USA\\
        \and
	   Department of Physics and Astronomy, University of Utah, Salt Lake City, UT 84112, USA\\
        \and
            Department of Astronomy, Nanjing University, Nanjing 210011, China (NJU)\\
   }

   \date{Received~~2012 month day; accepted~~2012~~month day}

\abstract{We describe a general target selection algorithm that is
  applicable to any survey in which the number of available candidates is
  much larger than the number of objects to be observed. This routine
  aims to achieve a balance between a smoothly-varying,
  well-understood selection function and the desire to preferentially select
  certain types of targets. Some target-selection examples are shown
  that illustrate different possibilities of emphasis
  functions. Although it is generally applicable, the algorithm was
  developed specifically for the LAMOST Experiment for Galactic
  Understanding and Exploration (LEGUE) survey that will be carried out using the Chinese Guo Shou Jing Telescope. In particular, this
  algorithm was designed for the portion of LEGUE targeting the
  Galactic halo, in which we attempt to balance a variety of science
  goals that require stars at fainter magnitudes than can be
  completely sampled by LAMOST. This algorithm has been implemented
  for the halo portion of the LAMOST pilot survey, which began in
  October 2011.
}

   \authorrunning{J.~L. Carlin, S.~L\'epine, \& H.~J. Newberg, et al. }            
   \titlerunning{LEGUE Target Selection Algorithm}  

   \maketitle

%
%
\section{Introduction}           
\label{sect:intro}

This document describes the algorithms used to select stellar targets for the
Milky Way structure survey known as LEGUE (LAMOST Experiment for
Galactic Understanding and Exploration). LEGUE is one component of the LAMOST Spectrocopic
Survey (see Zhao et al. 2012 for an overview) that will be carried out on the Chinese Guo Shou Jing Telescope (GSJT). The GSJT has a large (3.6-4.9-meter, depending on the direction of pointing) aperture and a focal plane populated with 4000 robotically-positioned fibers that feed 16 separate spectrographs, providing the opportunity to efficiently survey large sky areas to relatively faint magnitudes. 

The motivation for this algorithm was a desire for a
well-understood and reproducible selection function that will enable
statistical studies of Galactic structure. A continuous
selection function is desirable, rather than assigning targets by, for
example, ranges in photometric color, and excluding targets outside
the color-selection ranges. Another motivation for this scheme was the
opportunity provided by the sheer scale of the planned LAMOST survey;
the possibility of observing a large fraction of the available Galactic stars (at
high latitudes, at least) along any given line of sight allows for
less stringently-defined target categories, since a more general
selection scheme can gather (nearly) all of the stars in particular
target categories while simultaneously sampling all other regions of
parameter space. This opens up a large serendipitous discovery space
while also enabling studies of all components of the Milky Way. 

The LEGUE survey will obtain an unprecedented catalog of millions
of stellar spectra to relatively faint magnitudes (to at least 19th
magnitude in the SDSS $r$-band) covering a large contiguous area of
sky. The only large-scale spectroscopic survey of comparable depth is
the Sloan Extension for Galactic Understanding and Exploration (SEGUE;
Yanny et al. 2009a), which has been an enormously valuable resource for
studies of Milky Way structure (e.g., Allende Prieto et al. 2006, Carollo et al. 2007, Xue et al. 2008, Dierickx et al. 2010, Chen et al. 2011, Lee et al. 2011, Cheng et al. 2012, Smith et al. 2012) and substructure (e.g., Newberg et al. 2002, Yanny et al. 2003, Belokurov et al. 2006, Grillmair \& Dionatos 2006, Newberg et al. 2007, Klement et al. 2009, Schlaufman et al. 2009, Smith et al. 2009, Yanny et al. 2009b, Xue et al. 2011)\footnote{Note that these reference lists are meant only to give some representative Galactic (sub-)structure studies from SDSS, and are far from complete.}. However, the
SEGUE survey was limited to $\sim300,000$ stellar spectra in
$\sim600$ separate 7 square degree plates spread over the Sloan
Digital Sky Survey Data Release 8 (DR8; Aihara et al. 2011) footprint. The separation of SEGUE
into discrete "plates", while providing sparse coverage of all of the
Galactic components (as well as sampling a number of known
substructures), creates some difficulty in interpreting results from
SEGUE. In addition, the limited number of targets observed by SEGUE
necessitated selecting small numbers of stars from carefully defined
target categories, most of which were delineated by selections in
photometric color (Yanny et al. 2009a). This "patchy", non-uniform
selection function makes statistical studies of Galactic structures
difficult. The large contiguous sky coverage and sheer number of
targets that will be observed by LAMOST can help to overcome the
limitations of SEGUE for studies of Galactic structure; however, this
requires that the selection of LEGUE targets be done in a
well-understood, simply-defined manner.

Other spectroscopic surveys of large numbers of stars have focused on
magnitude-limited samples. For example, the Radial Velocity Experiment
(RAVE; Steinmetz et al. 2006), is a survey of $\sim1$~million stars to
limiting magnitude of $I = 12$ in the southern hemisphere. Upcoming
surveys, such as the HERMES Galactic Archaeology project (e.g., Barden
et al. 2008, Freeman \& Bland-Hawthorn 2008, Freeman 2010) will also observe magnitude-limited samples
of stars (in this case, to V=14). Obviously, a magnitude limited
survey does not require careful selection of subsets of available
targets, as is required for deeper surveys such as SEGUE or LAMOST.

In this paper, we present a general target selection algorithm
designed for surveys such as LAMOST where the number of available
candidates is much larger than the number of objects to be observed. The
method is sufficiently general to be extensible to any target selection
process, and can use any number of observables (i.e., photometry,
astrometry, etc.) to perform the selections. The paper is organized as
follows: we introduce a general target selection algorithm and show some examples of different selection biases that can be applied.
We follow this with a hypothetical survey design, discussing
the priorities for target selection in this mock survey, then show examples of the
adopted target selection parameters for a moderate latitude
($b\sim30^\circ$) and a high latitude ($b\sim60^\circ$) field. This hypothetical survey has target priorities similar to those outlined by Deng et al. (2012), based on the LEGUE science goals. More details of the use of our target selection algorithm for the LEGUE pilot survey can be found in Yang~et~al.~(2012), which discusses the dark nights portion of the pilot survey, and Zhang~et~al.~(2012), where a summary of the bright nights observing program is given (see also Chen et al. 2012 for discussion of an alternative target selection process that was applied to the Galactic disk portion of the LEGUE pilot survey). We follow the example survey illustration with some discussion about the difficulty in recreating a ``statistical sample'' of stellar populations from the observed set of spectra.
The
algorithms developed here have been used mostly with Sloan Digital Sky
Survey (SDSS) photometry as inputs (though see Zhang et al. 2012 for an example using 2MASS data), but in practice any
photometric, astrometric, spectroscopic, or other data known about the
input catalog stars can be used in the selection process. The target
selection programs discussed in this work were developed in the IDL
language.

\section{Target Selection Algorithms}

We initially set out to solve the general spectroscopic survey target
selection problem: starting with an input catalog of stars with any
number of ``observables'' (e.g., SDSS, with {\it ugriz} magnitudes,
positions, proper motions, etc.), define a general target selection
algorithm that is capable of producing the desired distribution of
targets. The assumption is that one begins with a large input catalog,
where the number of sources is larger than the number of objects that
can be observed with LAMOST. An input catalog would be a data table
with $N_{\rm S}$ stars for which we have $N_{\rm O}$ observables
(e.g., right ascension, declination, magnitude, color, proper motion
component, etc.):
\begin{equation}
\lambda = [\lambda_{\rm i}]_{\rm j}
\end{equation}
\noindent where $j=1,2,...,N_{\rm S}$ denotes any one of
$N_{\rm S}$ stars in the input catalog, and $i=1,2,...,N_{\rm O}$
denotes any of $N_{\rm O}$ observables which are available for every
star. To select targets for a spectroscopic survey such as LEGUE, one
would minimally require sky coordinates and a magnitude ($N_{\rm
  O}\geq3$).

For every LAMOST field, a number of stars can be randomly selected as
targets (based on how many fibers are available) among stars which are
located in the field, and for which each was assigned a statistical
weight. This statistical weight can be assigned according to a
function $P=P(\lambda_{1},\lambda_{2},...,\lambda_{N_{\rm O}})$
of the $N_{\rm O}$ observables, such that the probability for
selecting star $j$ as a target can be expressed as:
\begin{equation}
P_{\rm j} = P( [\lambda_{1}]_{\rm j}, [\lambda_{2}]_{\rm j}, ...,
[\lambda_{N_{\rm O}}]_{\rm j} )
\end{equation}
with the requirement
\begin{equation}
\sum_{\rm j}^{N_{\rm S}} P_{\rm j} = 1.
\end{equation}

\begin{figure}[!t]
\includegraphics[width=2.5in,trim=0.0in 0.0in 0.0in 0.0in,clip]{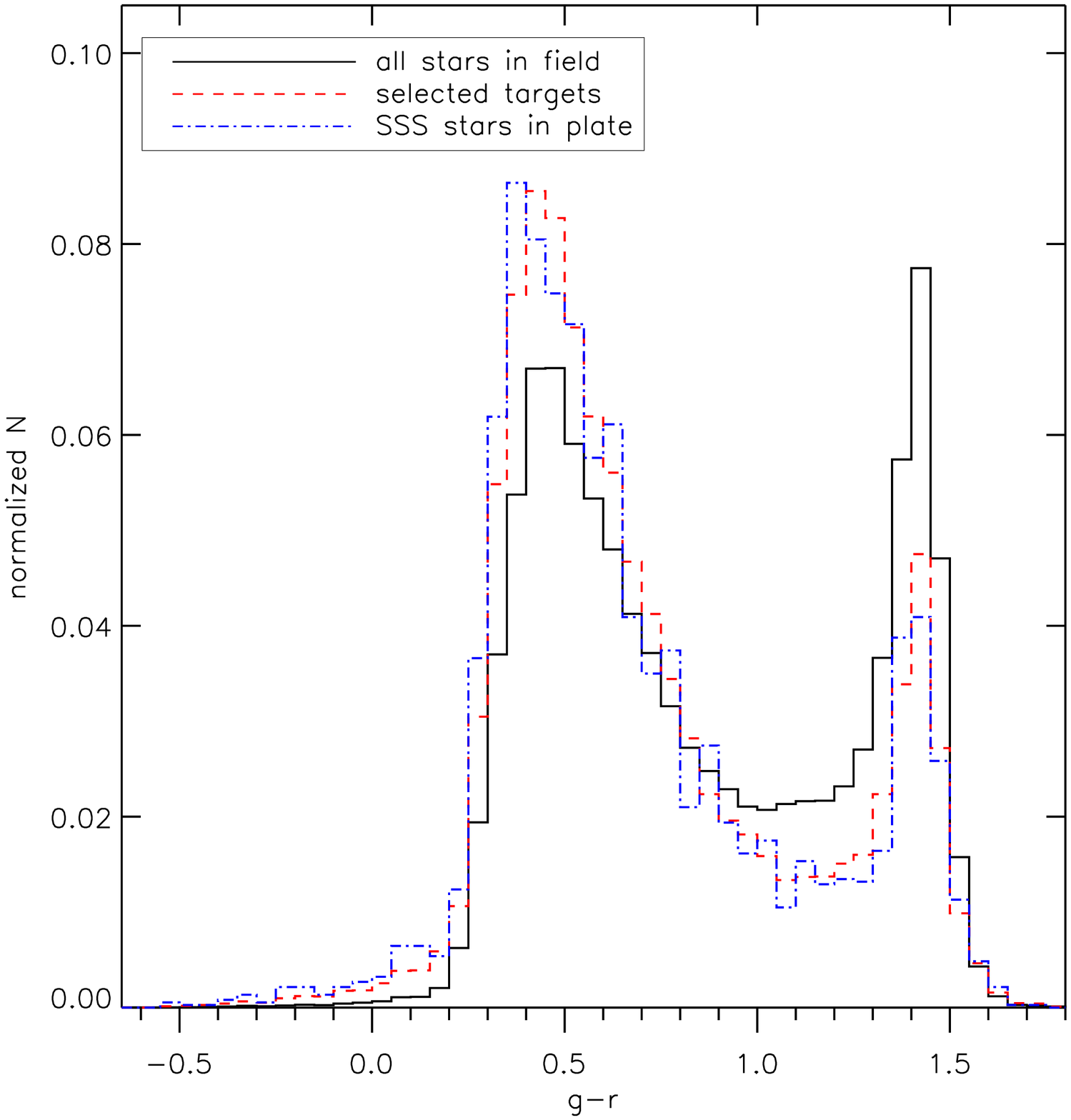}
\includegraphics[width=2.5in,trim=0.0in 0.0in 0.0in 0.0in,clip]{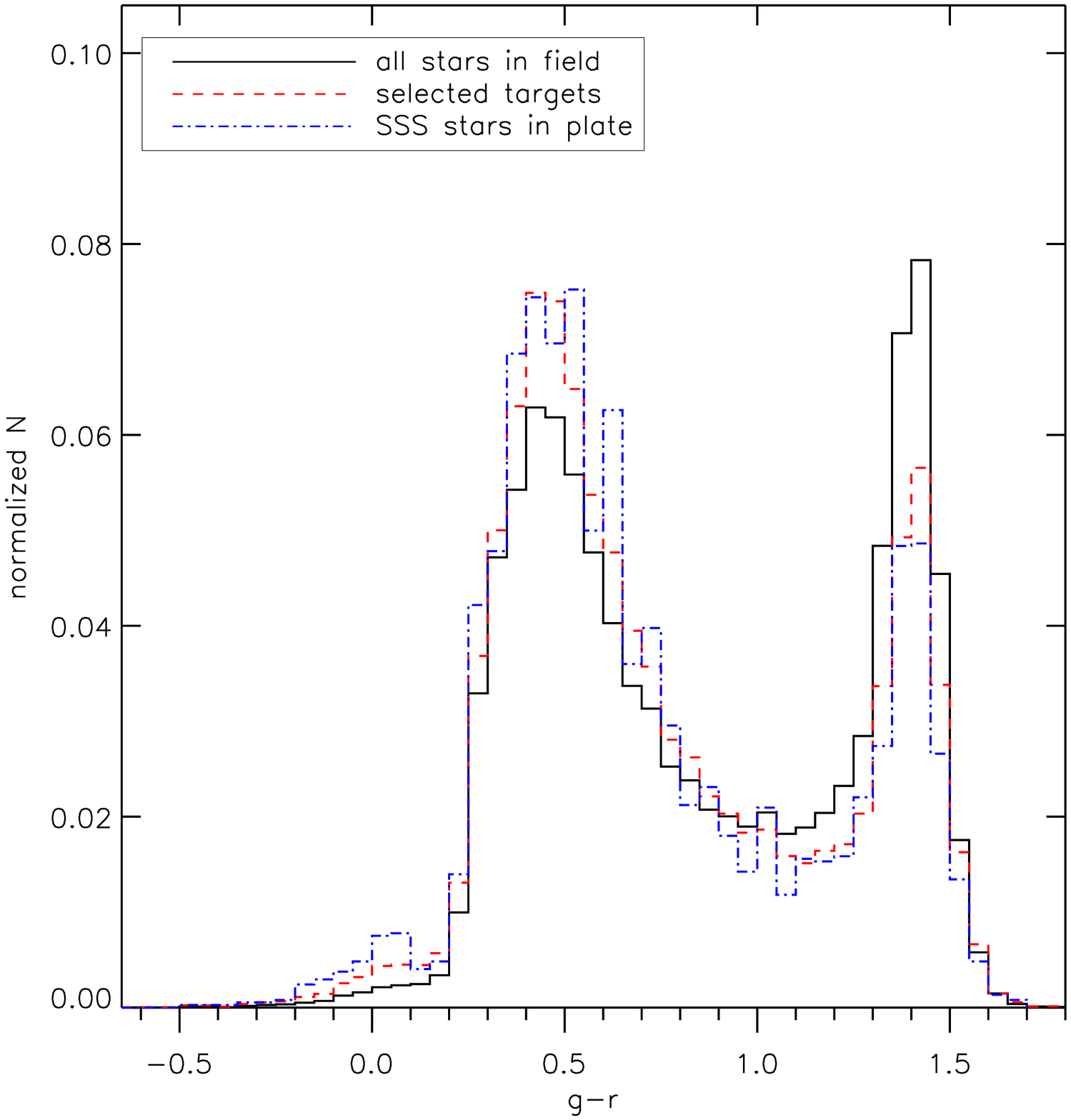}
\caption{Fractional distribution of $g-r$ color for stars in two
  sample regions of the sky $6^\circ \times 6^\circ$ in size
  (slightly larger than the area of a LAMOST plate); all data are
  selected from SDSS DR8. The left panel is for a field spanning
  $(\alpha, \delta) = (130-136^\circ, 0-6^\circ)$, corresponding to a
  field center in Galactic coordinates of $(l, b) \approx (225^\circ,
  28^\circ)$. This field contains a total of 102,199 stars between
  $14.0 < r < 19.5$. The right panel shows a field at $(\alpha,
  \delta) = (170-176^\circ, 0-6^\circ)$, or $(l, b) \approx
  (261^\circ, 59^\circ)$, containing a total of 44,566 stars. In both
  panels, the black solid line represents all stars in the field. The
  red dashed histogram is the stars selected by our algorithm as input 
  to the LAMOST fiber-assignment program (i.e., $\sim$600 per square
  degree), and the blue dot-dashed line shows the resulting
  distribution of spectroscopic targets in a single LAMOST plate
  (i.e., $\sim4000$ stars assigned to fibers). Each histogram has been
  normalized to 1, so that the bin heights represent the fraction of
  targets within each bin.
\label{fig:gr_hist}}
\end{figure}

The trivial case would be for every star to have the same probability
of being selected ($P_1=P_2=...=P_{\rm j}, \forall~j$). Calling this
trivial case ``model A'', and denoting its probability function
$P_{\rm j,A}$, then we have: 
\begin{equation}
P_{\rm j,A} = (N_{\rm S})^{-1} .
\end{equation}
Alternatively, one could base the selection on the statistical
distribution, $\Psi_0$, of the values taken by the observables. Defining
$\Psi_0$ as a continuous function over the $N_{\rm O}$ observables,
one would have:
\begin{equation}
\Psi_0 = \Psi_0( \lambda_{\rm 1}, \lambda_{\rm 2}, 
                 ..., \lambda_{N_{\rm O}} ) \equiv \Psi_0( \lambda_{\rm i} )
\end{equation}
\noindent which represents the density of recorded values for the
observables, normalized following
\begin{equation}
\int \Psi_0( \lambda_{\rm i})
  \prod_i^{N_{\rm O}} {\rm d}\lambda_{\rm i} = 1 .
\end{equation}

One way to determine $\Psi_0$ for the input catalog would be to
calculate the local density of sources at $(\lambda_1, \lambda_2, ...,
\lambda_i)$, estimated by counting the number of stars $j$ whose
observables satisfy the condition: 
\begin{equation}
\sqrt{\sum_i (\lambda_i-[\lambda_i]_j)^2} < \Delta\lambda
\end{equation}
where $\Delta\lambda$ defines the size of the volume in the space
of observables over which the stars are being counted, i.e., the
resolution of the function $\Psi_0$. For example, one could
determine the density function $\Psi_0=\Psi_0(g,g-r)$, calculating how
many stars can be found within 0.1 magnitudes of the parameter space
location $(g,g-r)$, which would mean using $\Delta (g,g-r)=0.1$
mag. This can be extended to any number of the observables to define a
"density" over multiple parameters; an example would be using
additional colors, calculating the number of stars within 0.1
magnitude of ($g, u-g, g-r, r-i$,...).

\begin{figure}[!t]
\includegraphics[width=2.5in,trim=0.0in 0.0in 0.0in 0.0in,clip]{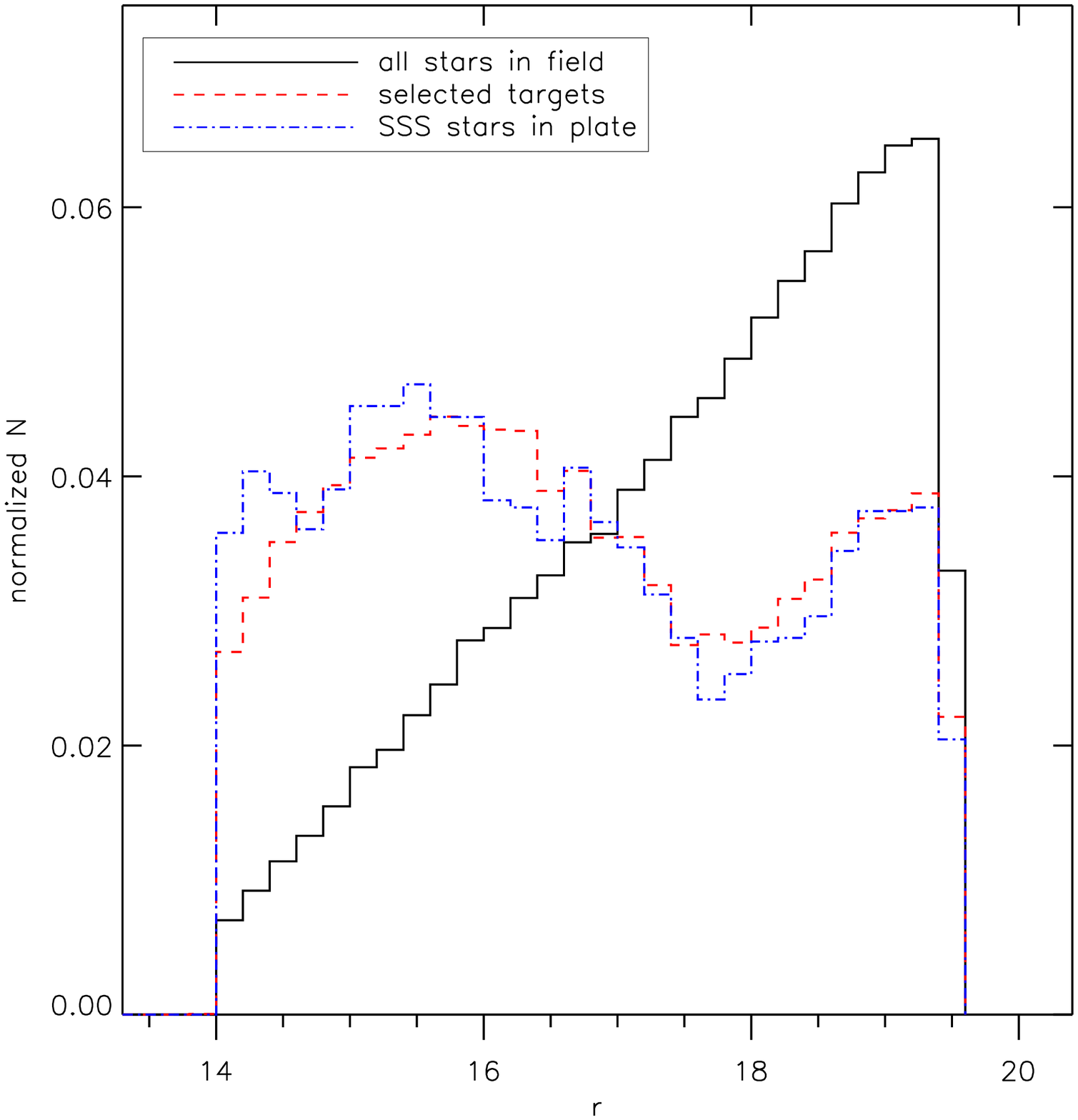}
\includegraphics[width=2.5in,trim=0.0in 0.0in 0.0in 0.0in,clip]{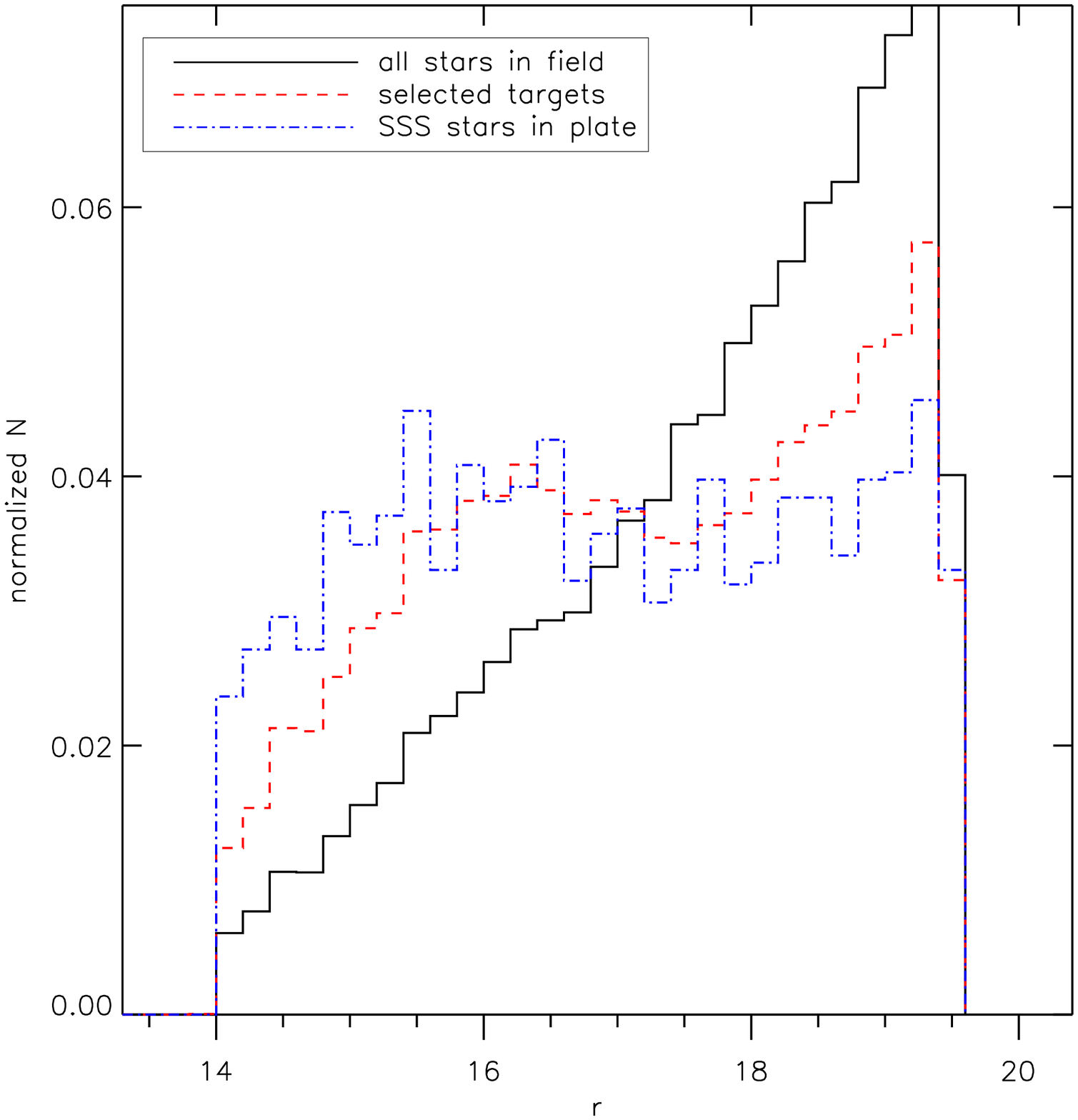}
\caption{Fractional distribution of $r$ magnitude for stars in the
  same two example regions as in Figure~\ref{fig:gr_hist}. The line
  styles and colors are also the same as in
  Figure~\ref{fig:gr_hist}. Each histogram has been normalized so that
  the bin heights represent the fraction of targets within each bin;
  this distribution can be thought of as a probability distribution,
  $\Psi_0 (r)$, of finding a star in each magnitude
  range. \label{fig:rmag_hist}}
\end{figure}

An example of a $\Psi_0$ function can be seen in
Figure~\ref{fig:gr_hist}, which shows the statistical distribution of
$g-r$ color for all stars in two sample
LAMOST plates as the solid black histograms. These have been
normalized so that the sum of all bins equals one, and can thus be
thought of as probability functions (in this case, $\Psi_0=\Psi_0
(g-r)$). A similar plot is seen in Figure~\ref{fig:rmag_hist} for $r$
magnitudes in the same two plates. For the remainder of this paper, we
will use examples from the simple case of the local density defined by
the number of stars within 0.1 magnitudes in $r$, $g-r$, and $r-i$,
i.e. $\Psi_0=\Psi_0 (r, g-r, r-i)$.

It is useful to point out that if one performs a target selection
following method ``A'', then the density distribution of the target
stars, denoted $\Psi_{\rm A}$, would be approximately the same as the
statistical distribution in the input catalog, to within the Poisson
errors, i.e. $\Psi_{\rm  A} (\lambda_{\rm i}) \approx \Psi_0
(\lambda_{\rm i})$.

\begin{figure}[!t]
\includegraphics[width=5.5in]{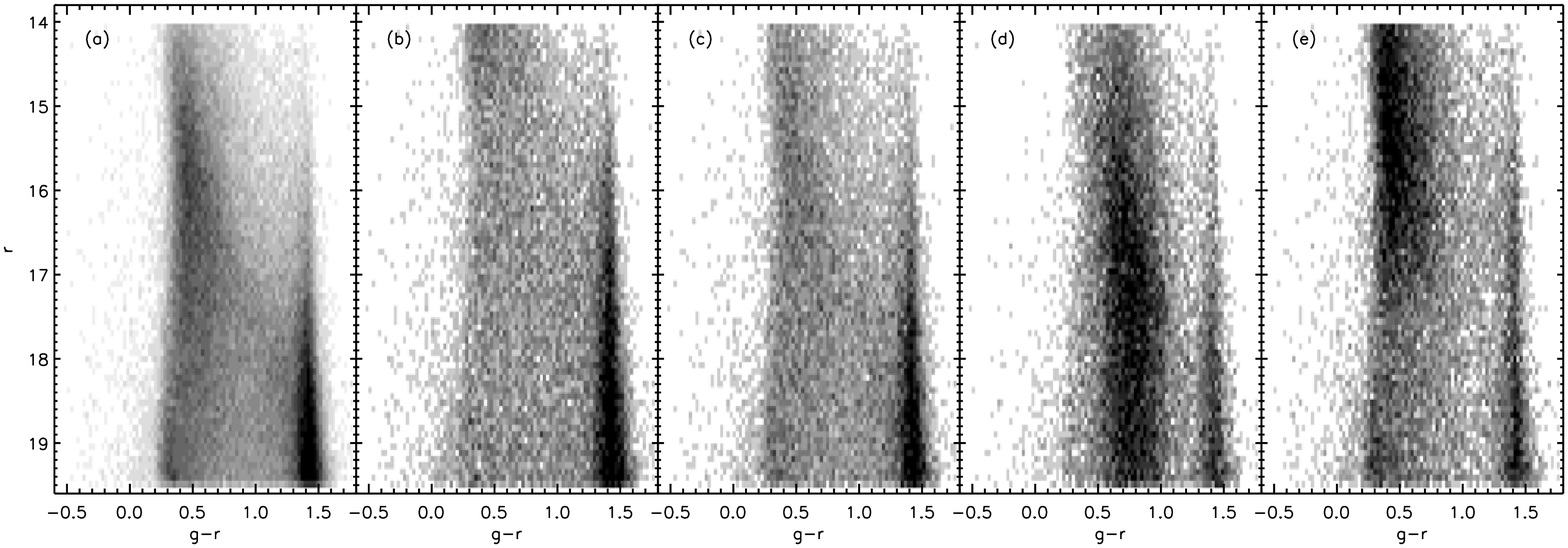}
\includegraphics[width=5.5in]{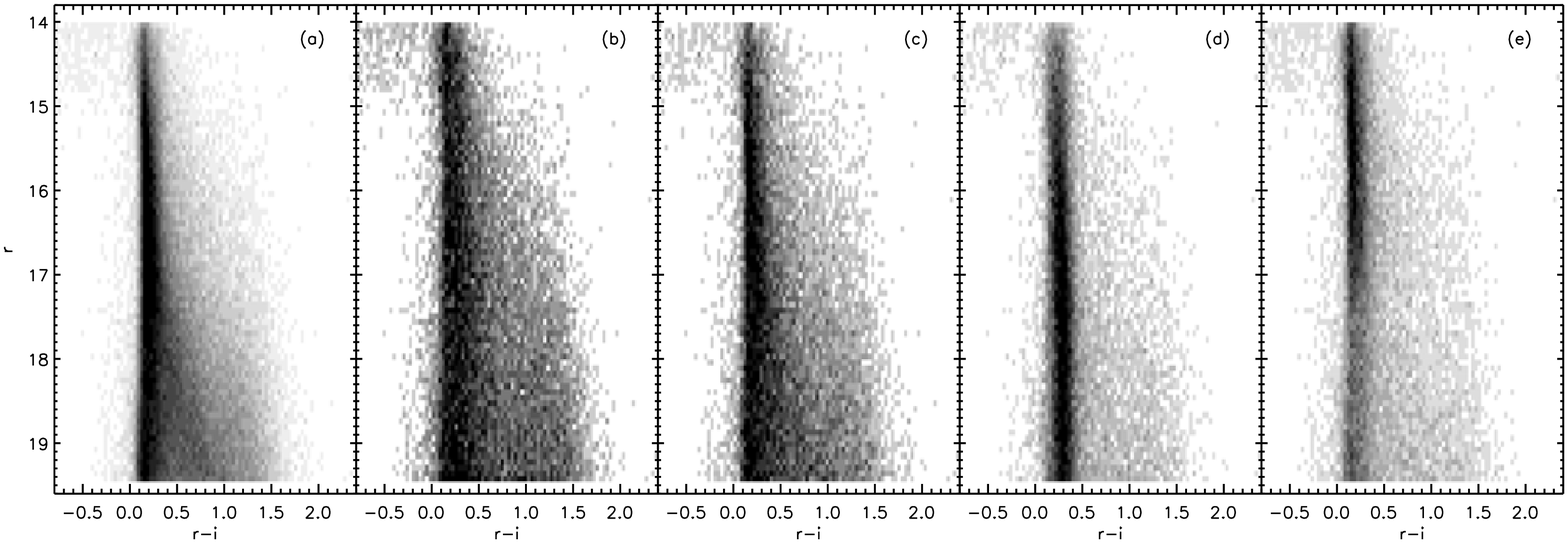}
\includegraphics[width=5.5in]{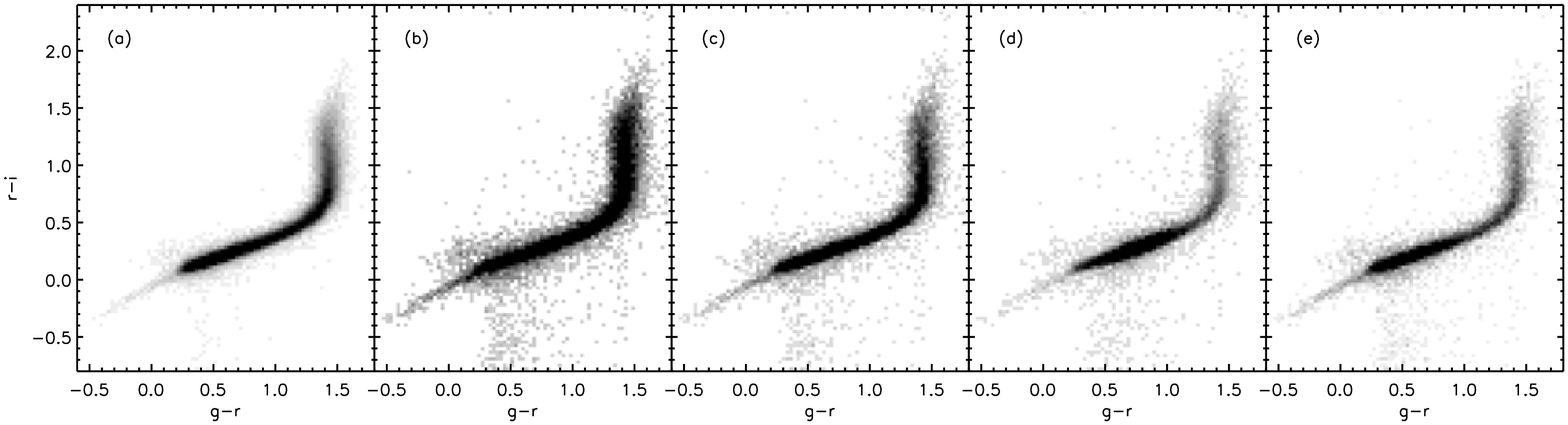}
\caption{Color-magnitude hess diagrams for different selections of
  stars in the region $(\alpha, \delta) = (130-136^\circ, 0-6^\circ)$,
  corresponding to a field center in Galactic coordinates of $(l, b)
  \approx (225^\circ, 28^\circ)$. Panel (a) shows all of the 102,199
  SDSS stars between $14.0 < r < 19.5$ in this field of view. Panels
  (b)-(e) show the results of selecting 600 stars per square degree in
  this field, with different selections (based on local density in
  three-dimensional $r$, $g-r$, $r-i$ space, or $\Psi_0 (r, g-r,
  r-i)$) represented in each panel. Panel (b) stars were selected with
  $\alpha = 1$, and panel (c) depicts the $\alpha = 0.5$ case. The $\alpha
  = 1$ selection (i.e., weighting by the inverse of $\Psi_0$) in panel
  (b) strongly de-emphasizes high-density regions of the CMD in favor
  of rare stars. M-stars at $g-r \sim 1.5$ appear oversampled in this figure; this arises because they are more spread out in $r-i$ colors than in $g-r$, causing them to be emphasized by the density weighting in 3-D parameter space. The overemphasis of rare objects is slightly less
  pronounced for $\alpha = 0.5$ (panel (c)), with a significant number of
  stars selected from the high-density regions of the CMD. In panel
  (d) we illustrate the results of selection with $\alpha = 0.5$ and with
  a over-selection of the region centered at $g-r = 0.8$ with width
  $\sigma_(g-r) = 0.2$ and overemphasis factor $A = 10$. This
  selection produces an overselection of stars centered at $g-r =
  0.8$, while retaining some stars from the remaining parameter
  space. Finally, panel (e) illustrates a selection with $\alpha =
  0.5$ and a linear bias in color beginning at $g-r = 1.1$ and sloping
  upward toward bluer colors with slope 2.5, and also a linear
  magnitude emphasis beginning at $r = 17.5$ with slope 1.0 toward
  brighter magnitudes. \label{fig:ra133_5panel}}
\end{figure}

\begin{figure}[!t]
\includegraphics[width=5.5in]{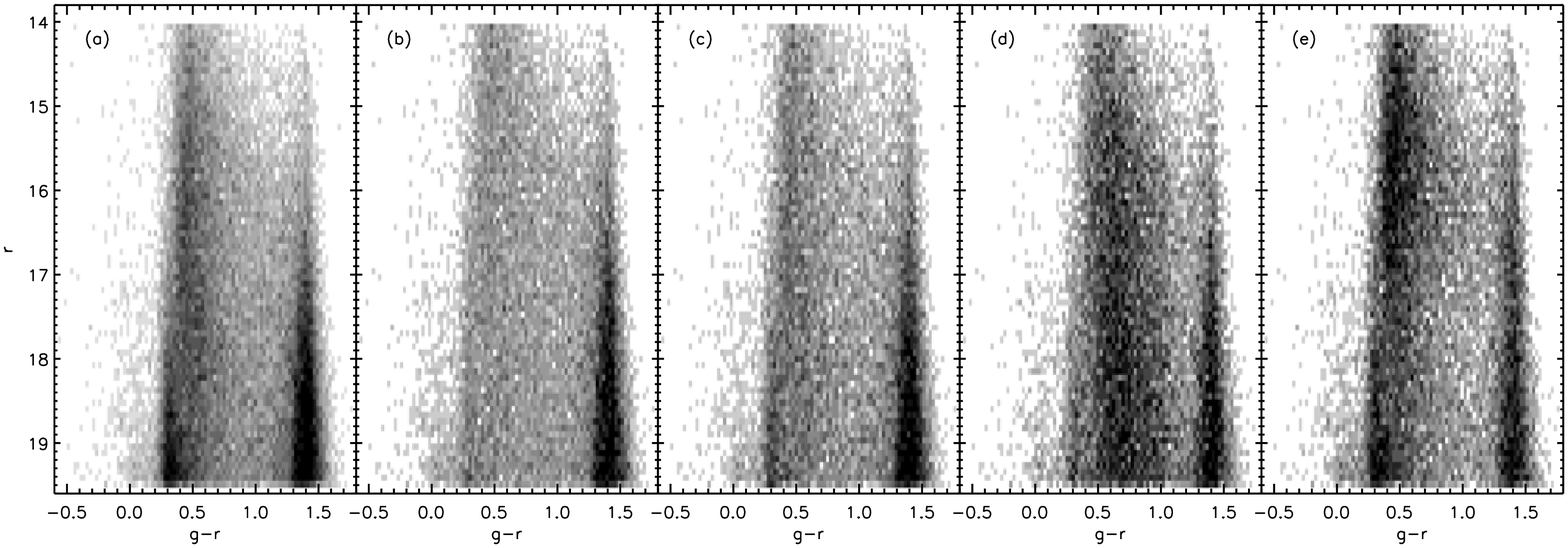}
\includegraphics[width=5.5in]{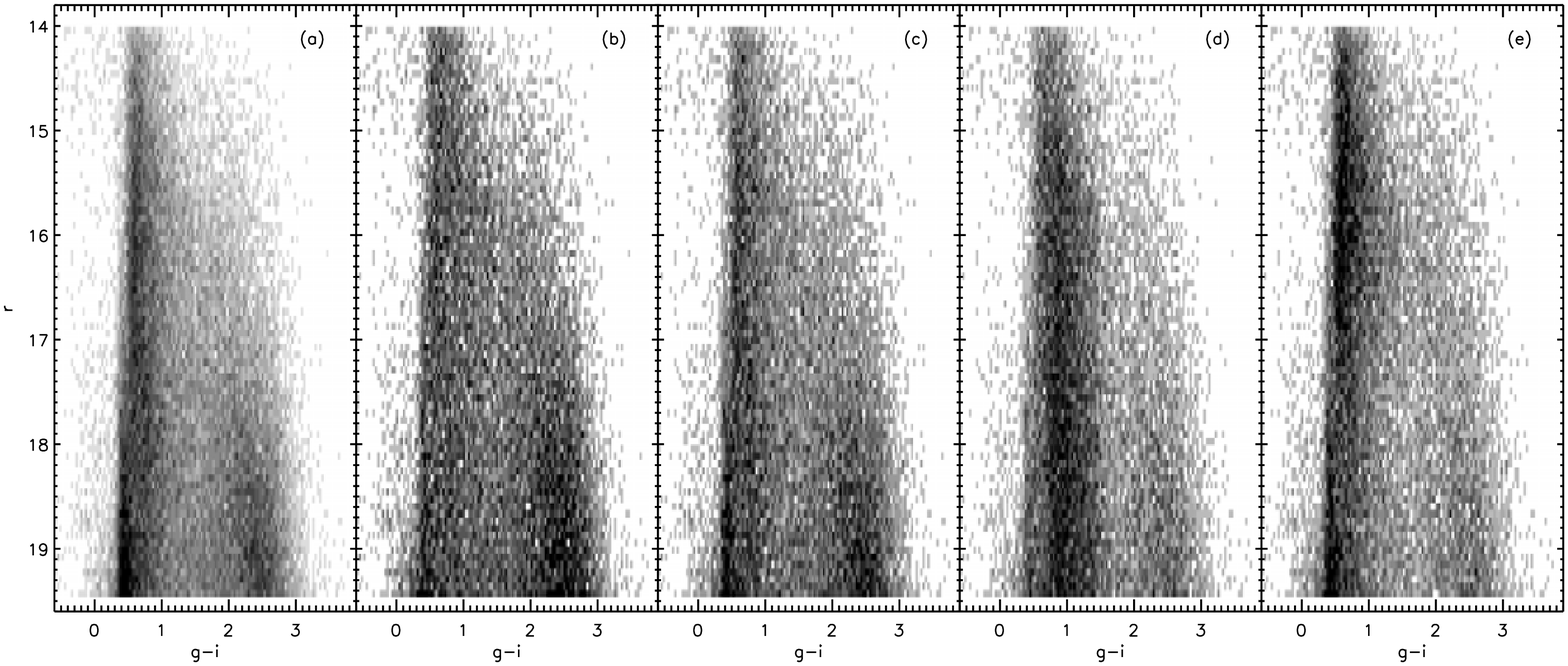}
\includegraphics[width=5.5in]{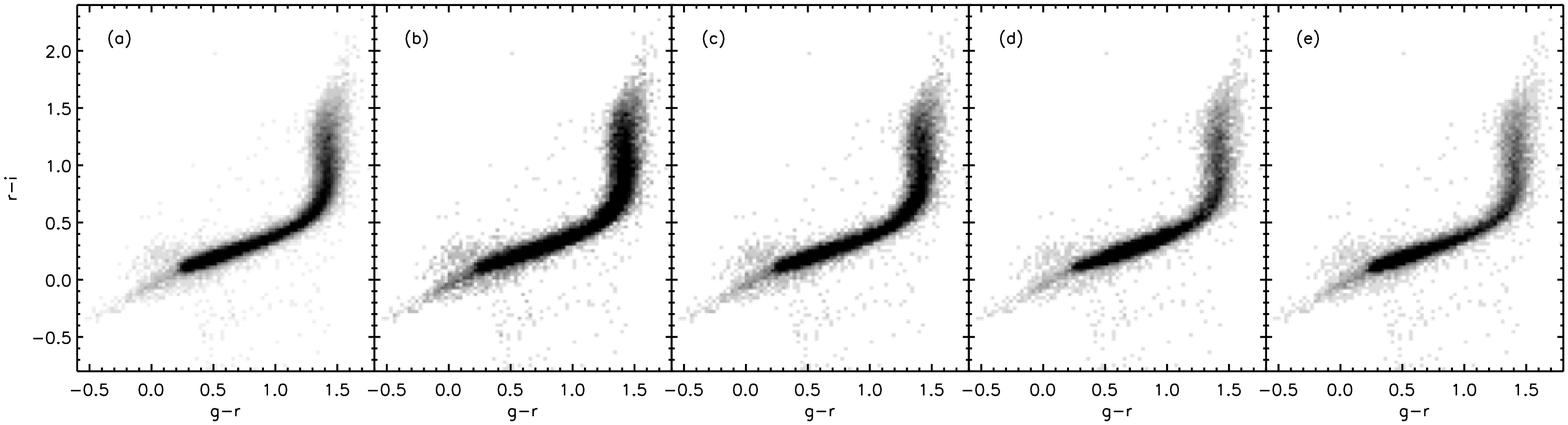}
\caption{As in Figure~\ref{fig:ra133_5panel}, but for a different
  field located at $(\alpha, \delta) = (170-176^\circ, 0-6^\circ)$, or
  $(l, b) \approx (261^\circ, 59^\circ)$. This higher-latitude field
  contains a total of 44,566 SDSS stars between $14.0 < r <
  19.5$. \label{fig:ra173_5panel}}
\end{figure}

Typically, however, one may want to obtain a list of targets whose
statistical distribution differs from the distribution in the input
catalog. For instance, one may want to overselect objects in a given
color/magnitude range, or pay more attention to outliers or unusual
stars. One possibility, which we will call "Method B", is to assign a
selection probability that is inversely proportional to the local
density in the space of observables:
\begin{equation}
P_{\rm j, B} = \frac{K_{\rm B}}{\Psi_0( [\lambda_{\rm 1}]_{\rm j},
  [\lambda_{\rm 2}]_{\rm j}, ..., [\lambda_{\rm N_o}]_{\rm j} )}
\end{equation}
\noindent where $K_{\rm B}$ is a normalization constant to ensure that
$\sum_{\rm j} P_{\rm j, B} = 1$. In Method B, the statistical
distribution of the selected stars ($\Psi_{\rm B}$) over the
observables ($\lambda_{\rm i}$) is different from that of the input
catalog ($\Psi_0$), and in fact it is to first order uniform over all
values of $\lambda_{\rm i}$, i.e.: $\Psi_{\rm B} (\lambda_{\rm i}) \approx
K_{\rm B}$. Examples of this type of selection are seen in panels (b) of Figures~\ref{fig:ra133_5panel} and \ref{fig:ra173_5panel}. Note, however, that because the local density was calculated using $r, g-r$, and $r-i$, the distribution does not look uniform in the color-magnitude diagrams (top and middle rows). However, the distribution in three-dimensional parameter space defined by $\lambda_{\rm i} = r, g-r, r-i$ should be roughly uniform.

As a generalization, one can assign probability that is inversely
proportional to some power of the local density, i.e.,
$[\Psi_0]^{-\alpha}$. In this case, which we will call Method C,
\begin{equation}
P_{\rm j, C} = \frac{K_{\rm C}}{[ \Psi_0( [\lambda_{\rm 1}]_{\rm j},
  [\lambda_{\rm 2}]_{\rm j}, ..., [\lambda_{\rm N_o}]_{\rm j} ) ]^{\alpha}}
\end{equation}
\noindent where $K_{\rm C}$ is a normalization constant to ensure that
$\sum_{\rm j} P_{\rm j, C} = 1$ . One can now see that Methods A and B represent
special cases where $\alpha = 0$ and $\alpha = 1$, respectively.
The random selection approach ($\alpha = 0$) would be ideal if one
simply wanted a selection that samples all of parameter space with the
same frequency as the input catalog. A weighting by $1/\Psi_0$ (i.e.,
$\alpha = 1$), on the other hand, produces an output catalog that
samples the space of observables more evenly, and thus contains a much
larger fraction of "rare" objects (i.e., those in lesser-populated
regions of parameter space, and de-emphasizing regions of higher local
density relative to the input catalog; see, e.g., panels (b) of
Figures~\ref{fig:ra133_5panel} and \ref{fig:ra173_5panel}). Adopting a
value $0<\alpha<1$ would result in a selection intermediate between
these two scenarios, i.e., one that increases the chances of rare
objects entering the selection, but still robustly samples the
high-density regions of the input distribution. The effects of using
$\alpha = 1/2$ are seen in panels (c) of Figures~\ref{fig:ra133_5panel}
and \ref{fig:ra173_5panel}. In the case where one would like to place
a particular emphasis on rare objects, a value of $\alpha>1$ could be
used, which would introduce a bias against the selection of more
common objects as targets. (Also, note that if there are fewer stars in a given region of parameter space than are selected in the more densely populated regions, that region will continue to be under-dense no matter what emphasis is applied.)

In any case, one may also be interested in over-selecting objects
occupying a particular region of parameter space (for example, a
narrow color or magnitude range, or simply selection of more blue than
red stars). To achieve this, an overemphasis can be included that favors the selection
of stars in a particular range of an observable by
including a bias explicitly in the selection probability
function. Any functional form of each of the observables, $\lambda_{\rm
  i}$, can be introduced to achieve the desired effect:
\begin{equation}
P_{\rm j, D} = \frac{{\prod}_i \ f_{\rm i} ([\lambda_{\rm i}]_{\rm j})}{[ \Psi_0( [\lambda_{\rm 1}]_{\rm j},
  [\lambda_{\rm 2}]_{\rm j}, ..., [\lambda_{\rm N_o}]_{\rm j} ) ]^{\alpha}}.
\end{equation}
Where the $f_{\rm i} (\lambda_{\rm i})$ can be any function of the
observables $\lambda_{\rm i}$. 

Two examples that are currently implemented for the LAMOST pilot survey are a local emphasis over
a specific range of colors, and a general bias over the magnitude range
to emphasize brighter stars or fainter stars. A local emphasis is
achieved using a function of the form:
\begin{equation}
f_{\rm i} (\lambda_{\rm i}) = 1 + A_{\rm i} e^{- \frac{(\lambda_{\rm i} - x_{\rm i})^2}{\sigma_{\rm i}^2}}
\end{equation}
\noindent where $x_{\rm i}$ is the central value of interest for the
observable $\lambda_{\rm i}$ (i.e., the center of the color or
magnitude range to emphasize), $\sigma_{\rm i}$ is the range of
interest, and $A_{\rm i}$ is the "over-selection" factor (i.e., how
strongly you wish to overemphasize these objects compared to stars
outside of this range). An example is shown in panels (d) of
Figures~\ref{fig:ra133_5panel} and \ref{fig:ra173_5panel}, where the
region on interest is centered at $g-r = 0.8$, with a range
$\sigma_{g-r} = 0.2$ and overemphasis factor $A = 10$.

Likewise, a general bias is introduced by the use of a linear
function of the form:
\begin{equation}
f_{\rm i} (\lambda_{\rm i}) = 1 - m_{\rm i} (X_{\rm i} - \lambda_{\rm i})
\end{equation}
\noindent where $m_{\rm i}$ is the slope of the linear emphasis
function, and $X_{\rm i}$ is the limiting value where the linear
emphasis ends (either the minimum or maximum allowed value of
$\lambda_{\rm i}$). This produces a function that is 1.0 at one
extremity, and increases to higher values from the limiting value
$X_{\rm i}$. An example of this would be to use a linear
function that increases toward lower values to overselect stars of
bluer colors and/or brighter magnitudes. The effects from this type
of selection bias are shown in panels (e) of
Figures~\ref{fig:ra133_5panel} and \ref{fig:ra173_5panel}; the color
selection is anchored at $g-r = 1.1$ with slope 2.5 increasing toward
bluer colors, and the magnitude emphasis has slope 1.0 anchored at $r
= 17.5$, increasing toward the bright end.

\subsection{Selecting Stars Using the Assignment Probabilities}

Here we describe the method we have implemented to select target stars based on the selection probability function (i.e., $P_{\rm j}$). Once the assignment probability for each star in the input catalog has
been defined, a cumulative probability is calculated for each star on
the list which consists of the sum of the probabilities of all stars in
the list up to (and including) that particular star:
\begin{equation}
P_{\rm cum, j} = \sum_{\rm k=0}^{\rm j} P_{\rm cum, k}
\end{equation}
\noindent where "j" is the index of each star. A random number is then
generated (using a uniform distribution from 0.0 to 1.0), and one 
star from the list is identified for which the random number is less than
the cumulative probability, but greater than the cumulative
probability of the previous element (i.e., where $P_{\rm cum, j-1} <$
randnum $< P_{\rm cum, j}$). This star is placed in the list of
"selected" stars and removed from the sorted list of candidates for
selection.
The assignment probabilities are then renormalized so that they
sum to unity once more, and the cumulative probabilities are
recalculated. (Note that the order the
stars appear in the input catalog does not matter, since a star with a
larger selection probability will carve out a larger range of the
cumulative probability space, and thus be more likely to be selected.)
The process is repeated until the desired number of stars has been selected. 
Selecting targets in this manner has the effect of preferentially choosing stars with higher selection probabilities, but still selecting some stars from the entire range of parameter space. This also means that the stars selected near the beginning will have different ``demographics'' (i.e., occupy different distributions in parameter space) than those selected later. Thus if a region of sky is revisited for a second (or more) observation, the distribution of targets in the observables will differ from the overall distribution in that same field of view. This in turn means that the selection probability as a function of the observables will also differ when revisiting a region of sky.

In the case of the LAMOST pilot survey, the number of targets to select is set by the
fiber-assignment software's requirement that the input catalog contain
roughly three times the desired number of spectroscopic targets. Thus,
since LAMOST contains 200 fibers per square degree in the focal plane,
we select 600 stars deg$^{-2}$ in the input catalogs (though this target
density is a parameter that can be set when running the program). This
is achieved by dividing the sky into $2\times2^\circ$ blocks, and
selecting targets in each block until the target density has been
reached.  Defining the local density separately for each of these
blocks has the benefit of mitigating the effects of large-scale
spatial variations of stellar populations within the survey footprint
on the defined local densities (and thus the target selection
probabilities). Note, however, that because three times the fiber density is required in the input catalog, only 1/3 of the targets selected for input to the fiber assignment algorithm will be observed. 
If the target selection was being done at the same time the fibers were being assigned to objects, one could maximize the probability that objects in a particular parameter range were selected if at all possible by assigning a very large probability in that range. Because the probabilities are pre-assigned separately from the fiber assignment process, we have implemented a priority scheme to preserve information about which objects would have been selected first. In the absence of this priority scheme, all objects sent to the fiber assignment algorithm would be observed with $\sim1/3$ probability, so it would be impossible to regularly observe more than 1/3 of any type of object.
Of course, revisiting the same plate multiple times increases the chance of observing all objects with certain selection criteria, since those that were unable to be assigned on the first plate can be picked up on later observations.

The LAMOST target assignment algorithm allows us to assign priorities from 0-99 for each of the selected objects, with lower numbers indicating higher priority for selection by a fiber.
The probability for selection calculated by the target selection algorithm must be converted to an integer priority value for the fiber assignment program, rather than being used directly to assign targets to fibers. 
When each fiber is being
assigned a target, all of the possible targets within its patrol radius are
examined, and the one with the lowest priority value assigned. If all
targets have been given equal priority, then the fiber will be
assigned to the target closest to its ``home" position; such an
instance would thus produce a uniform spatial distribution of targets,
ignoring any selection preferences based on photometry and other
properties. This also means that if multiple high-priority targets are within the patrol radius of single fiber, only one of them will be assigned to a fiber. To ensure that the desired target distribution in
parameter space is achieved, one would ideally assign as many priority values as possible, so that the probability distribution created by the target selection algorithm would be followed closely.
To do this, 
we assign priorities from 1-$M$ (where $M < 99$; for the pilot survey we used $M=80$, with the remaining priorities are reserved for other possible uses) to the 600
stars selected in each square degree of sky. The priorities are
assigned by taking the total number of targets desired in a given
region (in this case, 600 per square degree), dividing by $M$, then
looping from 1-$M$, assigning this number of targets to each priority
value. Since the probability weighting should preferentially select
targets of higher interest at the beginning of the selection process,
this method ensures that lower priority values (i.e., higher chance of
being assigned) are given predominantly to the objects with high
probability for assignment, with lower-probability stars mostly having
priorities that will make them less likely to be assigned. In
practice, this complicates the statistical understanding of the target
distribution, but our tests have shown that this method in effect
reproduces our desired target distributions.


\section{A Sample Hypothetical Survey}

\subsection{Survey Goals and Target Categories}

With a general target selection algorithm developed, we now
explore the question of what combinations of parameters can be used to achieve different target selection goals. In a non-magnitude-limited spectroscopic survey with multiple science goals (for example, LEGUE), there will typically be certain types of objects that are valued more than others. An example might be blue horizontal-branch (BHB) star candidates. These are an extremely valuable resource for Galactic structure studies because they are relatively rare, intrinsically bright (making them ideal probes of the distant Milky Way halo), trace metal-poor populations typical of the halo, easy to derive distances for, and occupy regions in photometric colors that are not confused with many other types of objects. So, for example, a study that is interested in BHB stars could try to simply select all stars with SDSS colors $(g-r) < 0.0$ and $(u-g)$ colors unlike those of QSOs. With our target selection algorithm, these objects could easily be preferentially selected without resorting to something like an abrupt cutoff at a certain photometric color. This can be achieved in one of two ways (or a combination of both): first, an emphasis on rare objects (BHB stars have relatively low densities in color-magnitude or color-color diagrams; see, e.g., Figure~\ref{fig:gr_hist} for an illustration of the paucity of such blue stars) can be achieved via weighting by the inverse of the local density in color/color/magnitude space (or, even better, weighting by $\alpha=1/2$), and secondly, by adding a linear emphasis that increases blueward of some cutoff color. Examples of the density weighting are shown in panels (b) and (c) of Figures~3 and 4, illustrating the effect of weighting by $\alpha=1$ and $\alpha=1/2$, respectively. The number of rare, blue objects is enhanced in these relative to the fraction of blue stars in the input catalog. An additional linear weighting can be applied; for example, one could choose to multiply the local density by a linear function beginning at $g-r = 0.3$ and increasing blueward. Both of these methods will increase the number of blue stars selected, while avoiding an abrupt cutoff to the selection at $g-r = 0.0$.

Within a given collaboration, there may be many science goals (e.g., see Deng et al. 2012 for a discussion of the LEGUE science aims). Balancing the need for a well-understood selection function with numerous target types is easily done with the method we have outlined. Here we create a hypothetical survey to use as an example. Our example survey (which happens to very closely resemble many of the LEGUE goals) aims to study the Galactic halo, while also sampling a large number of nearby stars of all types for studies of the Galactic disk. The survey will use only SDSS photometry for target selection, with no constraints on proper motions or other properties, selecting among stars with $14 < r < 19.5$. 

Studying the halo requires intrinsically bright, easily identified tracers such as BHB stars or K/M-giants to probe to large distances, and also a large sample of F-type turnoff stars at all magnitudes. F-type stars occupy a color range of relatively unambiguous luminosity classification (other than the occasional asymptotic giant-branch star). The BHB stars are very blue ($g-r < 0.0$), while K/M giants are red stars with $g-r > 1.0$. F-turnoff stars have colors of roughly $0.2 < g-r < 0.5$. BHB stars (and to a lesser extent, K/M giants) occupy a region of low stellar density in color-magnitude space, while F-turnoff stars are very common. This proposed survey would also like to obtain spectra of metal-poor M subdwarfs, which are not distinguishable in the $g-r$ band from all other M-dwarfs, but separate clearly in $r-i$ colors, while still sampling a large number of nearby M-dwarfs that can be used to probe local kinematics. There is also interest in following up interesting discoveries with high-resolution spectroscopy, so we wish to overemphasize bright stars within reach of echelle-resolution spectrographs. Finally, there is a desire to sample all stellar populations, but preferentially observe rare objects first, in order to open up the discovery space to rare (and perhaps previously unknown) stellar types. Briefly, then, the target selection categories are as follows:

\begin{itemize}
    \item Sample a larger fraction of ``rare'' stars than the ``less rare''. As shown in
    Section~2, this is exactly what is achieved by the use of local
    density weighting, $P_{\rm j} \propto [\Psi_0 (\lambda_{\rm
        i})^{-\alpha}]_{\rm j}$, in assigning selection probabilities
    to each star. In particular, we have shown examples of $\alpha =
    1$ and $\alpha = 1/2$; the $\alpha = 1$ case selects predominantly
    more rare objects (i.e., it under-emphasizes parameter spaces of
    high stellar density) than the $\alpha = 1/2$ density weighting.
 
      \item Select nearly all stars with $0.1 < (g-r) < 1.0$ and $r <
      17$ at high Galactic latitudes, and sub-sampled at
      $b<40^\circ$.

      \item Select nearly all stars with $g-r < 0.0$ and $u-g$ colors
      that are unlike those of quasars (i.e., BHB and blue straggler candidates).

      \item Select a significant fraction of the stars with $0.0 <
      (g-r) < 1.0$ and $17 < r < 19.5$ and $u-g$ colors that suggest
      they are not quasars. The bluer side of the color range should
      be selected with a probability about twice the redder side of
      the range to emphasize the F-type turnoff stars.

      \item Select a large number of M dwarfs at all magnitudes.
    
\end{itemize}    

We note that the above discussion refers to a survey with multiple visits to each sky position, which can
thus meet the requirements of nearly-complete samples of some subsets
of stellar types (for example, the very blue $g-r < 0$ stars). However, in practice not all of the high priority targets can be placed on one observation due to constraints on fiber positioning. Thus, for a survey where only a single visit to each sky area is planned,
these "requirements" should be considered to mean that one would like
as many as possible of the stars in these categories.

\subsection{Adopted Target Selection Parameters}

Through many tests, it was determined that the simplest combination of
parameters striking a good balance between all these desired
categories of targets is:
\begin{itemize}
  \item $\alpha = 0.5$, which weighs by the inverse square root of the
    local density.

  \item Linear ramp bias function in $g - r$, beginning at $g - r =
    1.1$ and increasing blueward with a slope of 2.5.

  \item Linear ramp bias function in $r$ magnitude, beginning at $r =
    17.5$, increasing toward brighter stars with slope of 1.0.
\end{itemize}

\noindent The particular values of these parameters (and particularly of $\alpha=0.5$) were determined somewhat subjectively based on visual examination of the selected targets and statistical distributions of targets from separate categories (as seen in Table~\ref{tab:frac_select}, which will be discussed in more detail below). Of course, detailed analysis could be done to optimize these target selection parameters if desired. However, in the case of a survey such as LAMOST, which will observe large numbers of stars with a variety of science goals, we simply select these parameters to produce input catalogs that are broadly consistent with the desired target distributions and sample all of parameter space to some extent.


\begin{figure}[!t]
\includegraphics[width=5.5in]{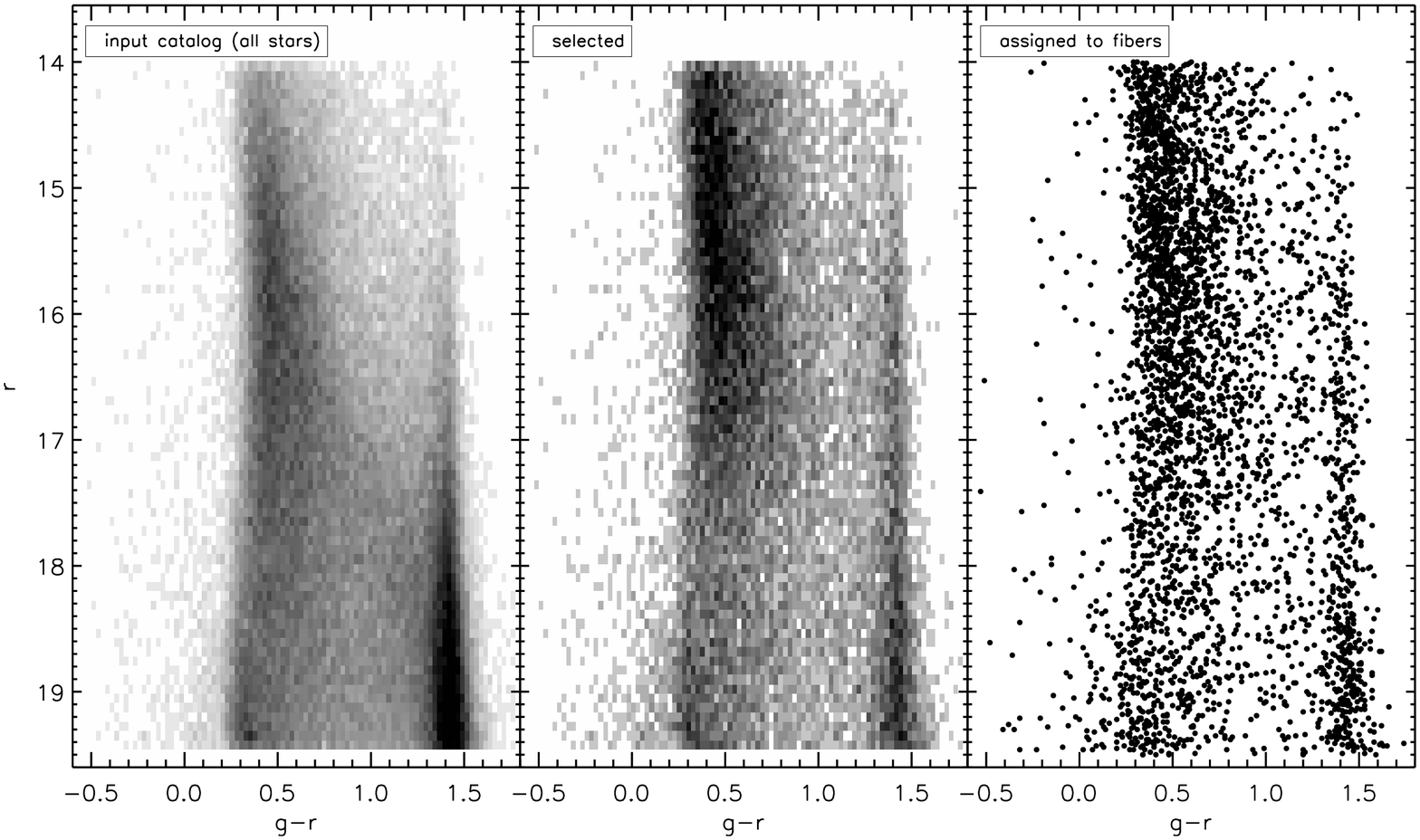}
\caption{{\it Left panel:} Color-magnitude hess diagram of all 112,099
  stars between $14 < r < 19.5$ selected from SDSS DR8 at
  $(\alpha,\delta)_{\rm J2000} = (130-136^\circ, 0-6^\circ)$. {\it
    Center panel:} Result of selecting 600 targets per square degree
  from the stars in the left panel. The target selection used $\alpha =
  0.5$ (to emphasize rare objects) with a ramp in color (to increase the
  fraction of blue stars selected) anchored at $g-r = 1.1$ and
  increasing blueward with slope of 2.5, and a ramp in magnitude (to
  weight bright stars more heavily) starting at $r = 17.5$ and
  increasing with slope of 1.0 toward the bright end. This selection
  represents 21.1\% of the stars in the field of view. {\it Right
    panel:} Distribution of targets assigned to LAMOST fibers upon
  running the catalog from the center panel through the
  fiber-assignment software. This panel contains a total of 3,715
  stars, or 3.6\% of the total number within the field of
  view. \label{fig:cmd133}}
\end{figure}

\subsection{Sample Target Selections}

\begin{figure}[!t]
\includegraphics[width=5.5in]{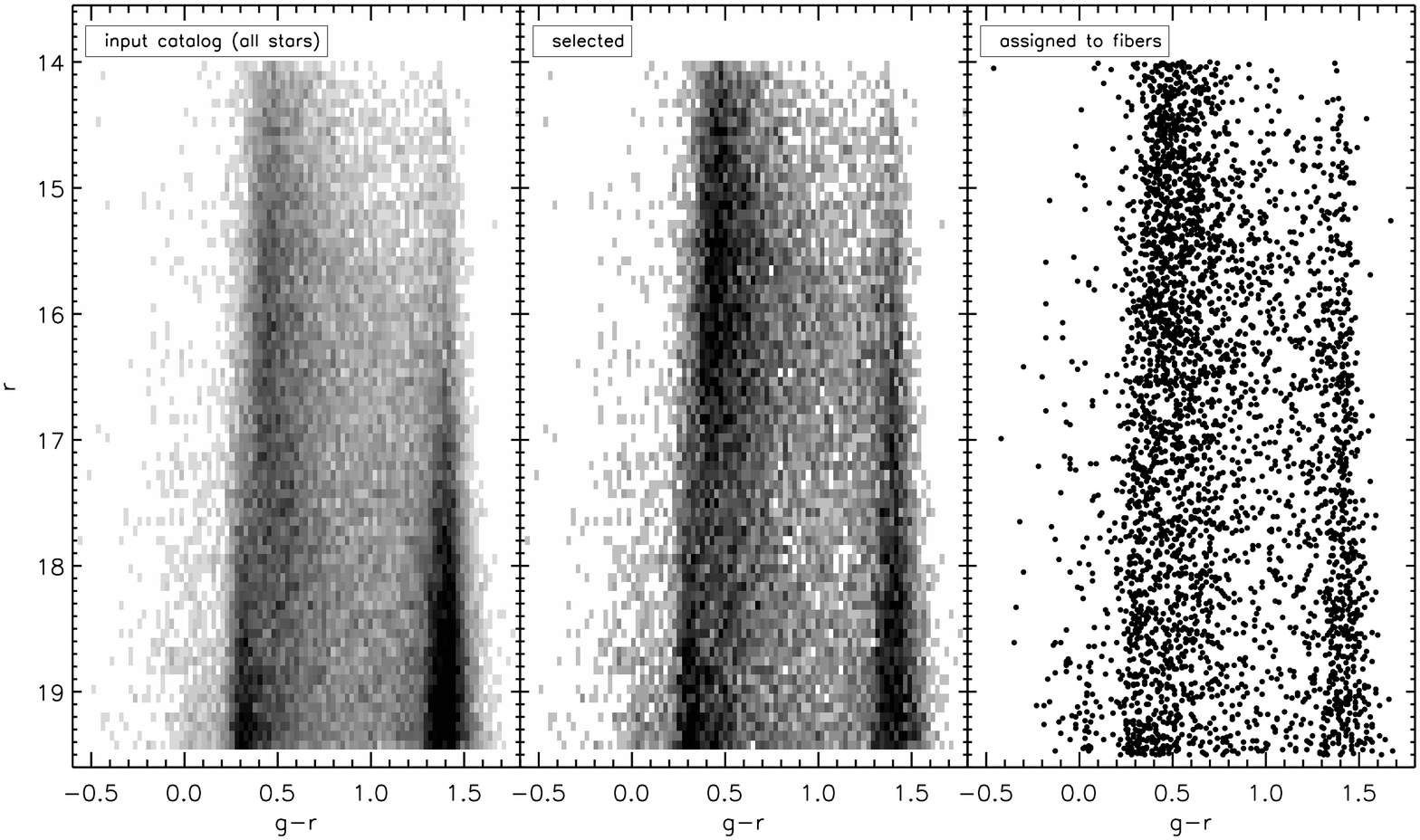}
\caption{As in Figure~\ref{fig:cmd133}, but for the higher-latitude
  field at $(\alpha,\delta)_{\rm J2000} = (170-176^\circ,
  0-6^\circ)$. This field of view contains a total of 44,566 stars, of
  which 48.4\% were selected for the center panel. Of these, 3722
  (seen in the right panel), or 8.4\%, were assigned to
  fibers. \label{fig:cmd173}}
\end{figure}

In this section, we show examples of outputs from the target selection
code, and follow this by selecting stars for targeting from among these using
the LAMOST fiber-assignment routine. These examples use the data from the same two fields 
presented earlier in this work, selected from $6\times6^\circ$ fields
at $(\alpha,\delta)_{\rm J2000} = (130-136^\circ, 0-6^\circ)$ and
$(\alpha,\delta)_{\rm J2000} = (170-176^\circ, 0-6^\circ)$. These
fields were chosen to show an example of a field at somewhat low
latitude ($b \sim 30^\circ$), and another at high latitude ($b \sim
60^\circ$). The local density, $\Psi_0$, is calculated in each of
these fields using $r$ magnitudes and $g-r$, $r-i$ colors. For
reference, the $r$ vs. $g-r$ color-magnitude distribution of all
102,199 stars in the $\alpha \sim 133^\circ$ field of view is seen in
the left panel of Figure~\ref{fig:cmd133}, and the 44,566 stars in the
lower-latitude $\alpha \sim 173^\circ$ field in
Figure~\ref{fig:cmd173} (note that these are the same as the left
panels in Figures~\ref{fig:ra133_5panel} and \ref{fig:ra173_5panel}).

The center panels of Figures~\ref{fig:cmd133} and \ref{fig:cmd173}
show the results of running the target selection routine on the input
catalogs, using $\alpha = 1/2$, a linear ramp in color, beginning at $g-r
= 1.1$ and rising blueward with slope 2.5, and a ramp in magnitude,
increasing from $r = 17.5$ with slope 1.0 toward bright stars. These
selections contain 600 stars per square degree, the required target
density to be input into the fiber assignment program. In the field at
$\alpha \sim 133^\circ$ (Figure~\ref{fig:cmd133}), 21.1\% of the
112,099 total stars in the region were selected as candidates, and in
the higher-latitude $\alpha \sim 173^\circ$ field
(Figure~\ref{fig:cmd173}) this number rises to 48.4\% of the total
available stars.

After running the catalogs selected for these two fields of view
through the LAMOST fiber assignment program, $\sim3700$ stars in each
of the two fields are allocated to fibers (the remaining fibers are to
be used for sky and other calibration purposes). The right panels of
Figures~\ref{fig:cmd133} and \ref{fig:cmd173} show the stars assigned
to fibers in these two fields. Generally, it is clear that quite a few
"rare" objects (for example, at intermediate colors of $0.8 < g-r <
1.2$, or bright M-star candidates at $g-r \sim 1.4$) are selected by
this method, but that densely-populated regions of color-magnitude
space are well-sampled, too. Note the fairly dramatic overemphasis of
bright ($r < 17$), blue ($g-r < 1.0$) stars achieved by the weighting
scheme.

\begin{table}[!t]
\caption{Fraction of stars from each target category assigned to fibers in a
  single LAMOST plate.}
\begin{center}
\begin{tabular}{|c|c|c|c|c|c|c|c|c|c|c|}

\hline
RA & Dec & l & b & total stars & assigned & \% assigned & very blue & bluish, bright & bluish, faint & red \\
($^\circ$) & ($^\circ$) & ($^\circ$) & ($^\circ$) &   &   & (\%) & (\%) & (\%) & (\%) & (\%) \\
\hline
133$\pm3$ & 3$\pm3$ & 225 & 28 & 102199 & 3715 & 3.6 & 22.6 & 6.7 & 2.7 & 2.1 \\
173$\pm$3 & 3$\pm$3 & 261 & 59 &  44566  & 3722 & 8.4 & 27.0 & 15.5 & 6.9 & 5.7 \\
\hline
\end{tabular}
\end{center}
\label{tab:frac_select}
\end{table}%

To assess how well the algorithm achieved the list of target selection goals outlined in Section~3.1, we select stars from the color and magnitude
ranges in which specific target-selection goals were focused, and
explore the relative emphasis or de-emphasis achieved by our code. The
degree of emphasis can be seen by comparing the fraction of stars
selected within a given target category to the fraction of the total
number of stars in the field. These results are given for the two
example fields in Table~\ref{tab:frac_select}. The table lists the
number of stars in each of the two fields of view ("total stars"),
followed by the number assigned to fibers for a single LAMOST plate
("assigned"), and the percentage of the total stars that were assigned
to be observed ("\% assigned", or "assigned"/"total stars"). The
following four columns represent the four categories outlined in
Section~3.1: "very blue" stars with $g-r < 0$, "bluish, bright" stars
with $0.0 < g-r < 1.0$ and $r < 17$, "bluish, faint" stars with $0.0 <
g-r < 1.0$ and $r > 17$, and "red" stars with $g-r > 1.0$. In each of
these columns, we provide the percentage of the total number of stars in
the field that satisfy those criteria that were assigned to a fiber on
the plate. This percentage can be compared to the "\% assigned" column
to see over- or under-emphasis; i.e., if the target selection was
uniform across color-magnitude space, one would expect roughly the
same fraction of stars to have been assigned in each category. Thus,
for the $\alpha = 133^\circ$ field, the fact that 22.6\% of the very
blue stars were assigned compared to 3.6\% overall means that the
"very blue" stars have been overemphasized by a factor of
$>6$. Examination of Table~\ref{tab:frac_select} shows that, at least
broadly, we have achieved our goals of strongly over-selecting very
blue objects, increasing the fraction of bright, blue stars that gets
observed, yet still retaining a significant number of faint, bluish
stars and red K- and M-star candidates (note that $>800$ red stars
with $g-r > 1.0$ were assigned in each plate -- even though they have
been underemphasized, they are still well-represented).

\begin{figure}[!t]
\includegraphics[width=5.5in]{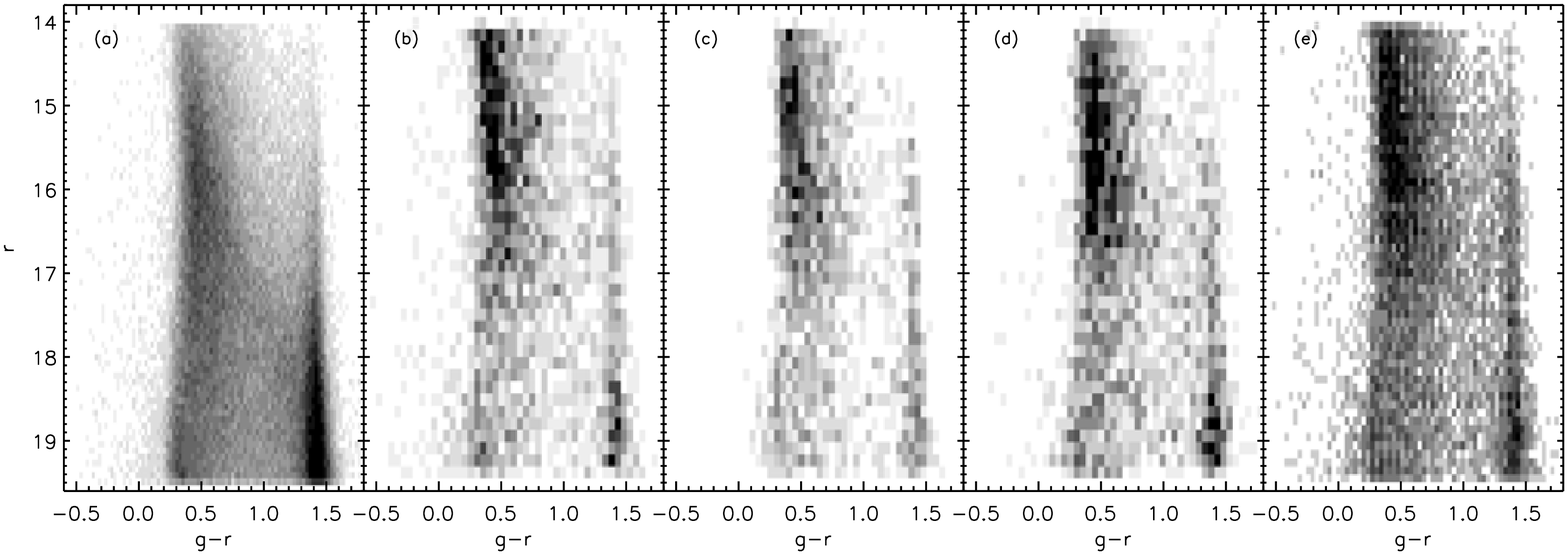}
\includegraphics[width=5.5in]{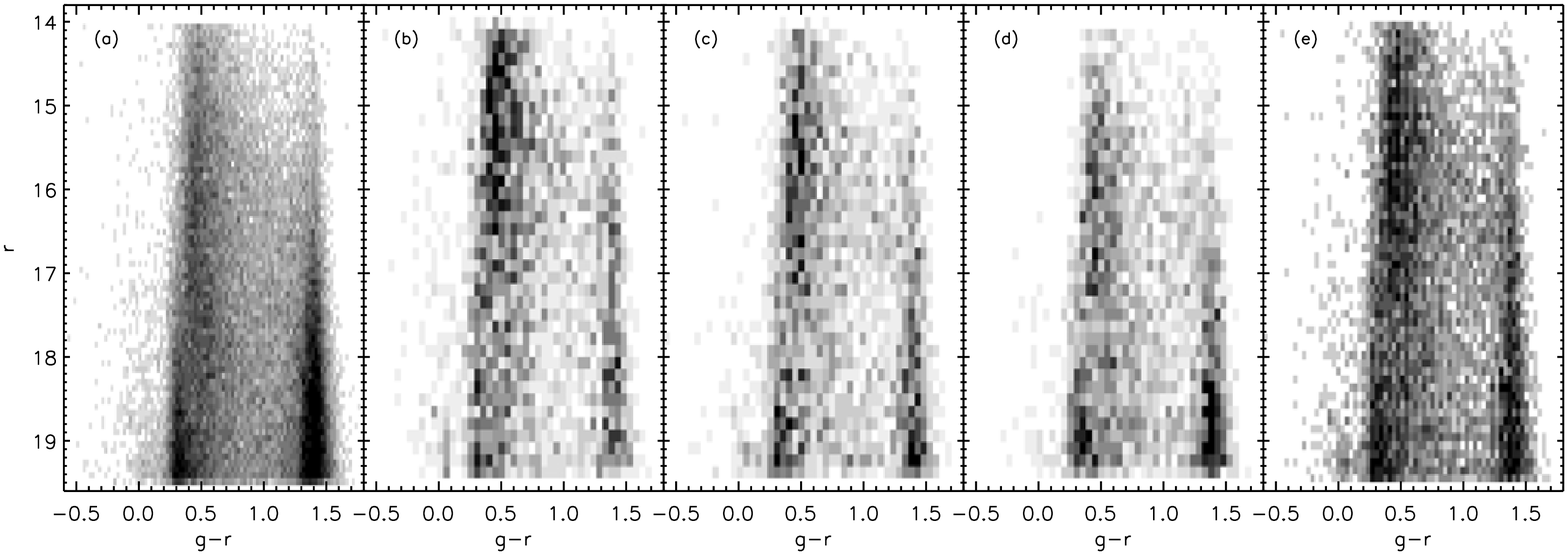}
\caption{Color-magnitude Hess diagrams for stars selected from the low-latitude ($b \sim 30^\circ$) field (upper panels) and high-latitude ($b \sim 60^\circ$) example fields. The same target selection parameters were used as in Figures~\ref{fig:cmd133} and \ref{fig:cmd173}. In each row, the panels represent (a) all stars from SDSS in the field of view, (b) stars selected for fiber assignment on the first LAMOST plate in this field, (c) a second plate excluding the stars in the first assignment, (d) a third plate, excluding stars from the first two, and (e) the sum of the three selected plates.
The high-latitude field of view contains a total of 44,566 stars, and the lower-latitude field has 102,199. Roughly 3700 stars were assigned on each of the three plates, so that in total, 24.4\% of the high-latitude stars were assigned to a fiber on one of the three plates, and 10.8\% of those at lower latitudes. In the low-latitude field there are many stars available, so the effects (especially the emphasis on bright, blue stars) of the preferential targeting categories are obvious in all three plates. However, the sum of all three plates (upper panel (e)) contains representative samples from all regions of color-magnitude space. The higher-latitude field (lower panels) has much lower stellar density. The first high-latitude plate (lower panel (b)) appears very similar to the corresponding selection from the low-latitude field, with bright, blue stars overemphasized (also note that many of the very blue, $g-r < 0.2$ objects are gone after the first plate). By the second and third plates in this field the selected stars start to cover more of the parameter space; once a large fraction of the bright, blue stars have been assigned, many more faint, red M-type stars get selected. 
\label{fig:3plate_cmds}}
\end{figure}

\section{Some Caveats About Statistical Target Selection}

Ostensibly, one of the reasons for having a smoothly-varying, well understood selection function is to be able to infer the underlying stellar populations from a given set of spectroscopically observed stars. However, in order for this to be possible, detailed records of the entire target selection and fiber assignment process need to be kept. The first issue affecting this is the need to supply the fiber assignment routine with a catalog with higher target density than the fiber density on the sky. Because of this, not all high priority targets will be placed on fibers. Some fibers in each observation will inevitably fail to yield useful spectra, making it necessary to factor the ``missed targets'' into analysis. 

Of course, any routine that preferentially targets certain objects will produce different target demographics if multiple visits to the same sky position are desired. An illustration of this is seen in Figure~\ref{fig:3plate_cmds}, which shows examples of 3 LAMOST plates selected in each of the two example fields used throughout this work. The upper panels show the relatively low-latitude ($b \sim 30^\circ$) field, and the lower panels show the $b \sim 60^\circ$ field. The target selection parameters were the same as those used in Sections 3.2 and 3.3, and seen in Figures~\ref{fig:cmd133} and \ref{fig:cmd173}. In each row, the five panels show $r$ vs. $g-r$ Hess diagrams of (a) the input sky distribution from SDSS, (b) the first plate selected, (c) the second plate selected (excluding stars from plate 1), (d) the third plate selected (excluding stars from plates 1 and 2), and (e) the sum of all three plates from (b)-(d). In the low-latitude field there are many stars available, so that the effects of the preferential targeting categories are seen in all three plates (especially the emphasis on bright, blue stars). However, the sum of all three plates (panel (e)) contains representative samples from all regions of color-magnitude space. The higher-latitude field (lower panels) is quite different. The stellar density is much lower in this field, so that in three LAMOST plates, a total of 24.4\% of the stars between $14 < r < 19.5$ are assigned to fibers. The first plate (panel (b)) appears very similar to the corresponding selection from the low-latitude field, with bright, blue stars overemphasized (note also that many of the very blue, $g-r < 0.2$ objects are gone after the first plate). By the 2nd and 3rd plates in this field, however, a large fraction of the bright, blue stars have already been assigned, and the selected stars start to cover more of the parameter space (specifically, there are many more faint stars -- especially a lot more faint, red M-type stars). Thus if one pre-selected three plates in a high-latitude field, the demographics of the stars on each observed plate would be quite different from each other. This makes reconstruction of the underlying populations rather difficult, because a different fraction of stars from each region of parameter space will have been observed depending on the stellar populations and stellar density in each field. Of course, variations in the number of times a given piece of sky is covered will dramatically alter the distribution of objects in the final catalog. 

Finally, we note that a routine that weights stars for selection based on the local density in parameter space will produce catalogs with different target demographics for different regions of sky. This is inevitable, because as we just showed, the stellar density on the sky actually affects the distribution of selected stars in parameter space, such that there is no way to get identical samples from regions of sky with different stellar densities. We also note that for a survey such as LAMOST, with a circular field of view, it is not possible to cover the whole sky with each part sampled only once. This will inevitably make the sampling of certain regions of sky higher than others.

Thus, to determine the underlying stellar populations based on the observed spectroscopic sample, one would need to either simulate the entire selection process, or compare the number of spectra of each type observed in a given part of the sky to the number of that same type of star that was available in the photometric catalog. We note that holistic models of the Galaxy with tuneable analytic parameters are now available (e.g., the Galaxia code; Sharma et al. 2011) which could be sampled with the selection function of the survey and used to correct survey artifacts.

\section{Conclusion}

We have presented a general target selection algorithm that can be
used in any instance where large numbers of stars are to be selected
from a catalog that is much larger than the desired number of
targets. The program performs selections in multi-dimensional
parameter space defined by any number of observables (or combinations
of observables). Various functions are available to emphasize certain
types of targets, and the program can be readily modified to implement
an overemphasis based on any smooth function of the observables. This
target selection algorithm was developed for the LEGUE 
portion of the LAMOST survey, and has been implemented in the LAMOST
pilot survey. We have shown that careful selection of the target selection parameters 
can produce the desired relative numbers of various target categories,
while retaining a smooth distribution across parameter space.

\begin{acknowledgements}
We thank the referee, Sanjib Sharma, for helpful comments on the manuscript. This work was supported by the US National Science Foundation, through grant AST-09-37523, and the Chinese National Natural Science Foundation (NSFC) through grant No. 10573022, 10973015 and 11061120454, and CAS grant GJHZ200812. S.~L. is supported by  U.S. National Science Foundation grant AST-09-08419. \end{acknowledgements}

\label{lastpage}

\end{document}